\documentclass[11pt,a4paper]{article}
\usepackage{cite}

\usepackage{amsmath, amsthm,mathtools, url,amssymb,upgreek,slashed}
\usepackage{ifpdf}
\ifpdf
  \usepackage[pdftex]{graphicx}
  \usepackage{epstopdf}
\else
  \usepackage[dvips]{graphicx}
\fi
\parskip=\baselineskip

\begin{document}
\begin{titlepage}
\begin{flushright}

\end{flushright}

\vskip 1.5in
\begin{center}
{\bf\Large{Deformations of JT Gravity and Phase Transitions}}
\vskip
0.5cm { Edward Witten} \vskip 0.05in {\small{ \textit{Institute for Advanced Study}\vskip -.4cm
{\textit{Einstein Drive, Princeton, NJ 08540 USA}}}
}
\end{center}
\vskip 0.5in
\baselineskip 16pt
\begin{abstract}  We re-examine the black hole solutions in classical theories of dilaton gravity in two dimensions.
We consider an arbitrary dilaton potential such that there are black hole solutions asymptotic at infinity to the nearly
$\mathrm{AdS}_2$ solutions of JT gravity, and such that the black hole energy and entropy are bounded below.
We show that if there is a black hole solution with negative specific heat at some temperature $T$, then at the same
temperature there is a black hole solution with lower free energy and positive specific heat.   As the temperature is
increased from 0 to infinity, the black hole energy and entropy increase monotonically but not necessarily continuously;
there can be first order phase transitions, similar to the Hawking-Page transition.  These theories can also have solutions corresponding
to closed universes.   \end{abstract}
\date{May, 2020}
\end{titlepage}
\def\SO{{\mathrm{SO}}}
\def\G{{\text{\sf G}}}
\def\la{\langle}
\def\ra{\rangle}
\def\g{{\cmmib g}}
\def\U{{\mathrm U}}
\def\M{{\mathcal M}}
\def\d{{\mathrm d}}
\def\CS{{\mathrm{CS}}}
\def\Z{{\Bbb Z}}
\def\R{{\Bbb R}}
\def\J{{\mathcal J}}
\def\Bbb{\mathbb}
\def\Tr{{\rm Tr}}
\def\16{{\bf 16}}
\def\1{{(1)}}
\def\2{{(2)}}
\def\3{{\bf 3}}
\def\4{{\bf 4}}
\def\sg{{\mathrm g}}
\def\i{{\mathrm i}}
\def\h{\widehat}
\def\u{u}
\def\D{D}
\def\sp{{\sigma}}
\def\E{{\mathcal E}}
\def\O{{\mathcal O}}
\def\PF{{\mathit{P}\negthinspace\mathit{F}}}
\def\tr{{\mathrm{tr}}}
\def\be{\begin{equation}}
\def\ee{\end{equation}}
 \def\Sp{{\mathrm{Sp}}}
 \def\Spin{{\mathrm{Spin}}}
 \def\SL{{\mathrm{SL}}}
 \def\SU{{\mathrm{SU}}}
 \def\SO{{\mathrm{SO}}}
 \def\ll{\langle\langle}
\def\rr{\rangle\rangle}
\def\la{\langle}
\def\ra{\rangle}
\def\T{{\mathcal T}}
\def\V{{\mathcal V}}
\def\bar{\overline}
\def\v{v}

\def\tilde{\widetilde}
\def\t{\widetilde}
\def\R{{\Bbb{R}}}\def\Z{{\Bbb{Z}}}
\def\N{{\mathcal N}}
\def\B{{\mathcal B}}
\def\H{{\mathcal H}}
\def\hat{\widehat}
\def\Pf{{\mathrm{Pf}}}
\def\PSL{{\mathrm{PSL}}}
\def\Im{{\mathrm{Im}}}
\font\teneurm=eurm10 \font\seveneurm=eurm7 \font\fiveeurm=eurm5
\newfam\eurmfam
\textfont\eurmfam=\teneurm \scriptfont\eurmfam=\seveneurm
\scriptscriptfont\eurmfam=\fiveeurm
\def\eurm#1{{\fam\eurmfam\relax#1}}
\font\teneusm=eusm10 \font\seveneusm=eusm7 \font\fiveeusm=eusm5
\newfam\eusmfam
\textfont\eusmfam=\teneusm \scriptfont\eusmfam=\seveneusm
\scriptscriptfont\eusmfam=\fiveeusm
\def\eusm#1{{\fam\eusmfam\relax#1}}
\font\tencmmib=cmmib10 \skewchar\tencmmib='177
\font\sevencmmib=cmmib7 \skewchar\sevencmmib='177
\font\fivecmmib=cmmib5 \skewchar\fivecmmib='177
\newfam\cmmibfam
\textfont\cmmibfam=\tencmmib \scriptfont\cmmibfam=\sevencmmib
\scriptscriptfont\cmmibfam=\fivecmmib
\def\cmmib#1{{\fam\cmmibfam\relax#1}}
\numberwithin{equation}{section}
\def\a{{\eusm A}}
\def\b{{\eusm B}}
\def\neg{\negthinspace}
\def\d{\mathrm d}
\def\C{{\Bbb C}}
\def\HH{{\mathbb H}}
\def\P{{\mathcal P}}
\def\NS{{\sf{NS}}}
\def\Ra{{\sf{R}}}
\def\sV{{\sf V}}
\def\Z{{\Bbb Z}}
\def\A{{\eusm A}}
\def\B{{\eusm B}}
\def\S{{\mathcal S}}
\def\bar{\overline}
\def\sc{{\mathrm{sc}}}
\def\Max{{\mathrm{Max}}}
\def\CS{{\mathrm{CS}}}
\def\ga{\gamma}
\def\bg{\bar\ga}
\def\W{{\mathcal W}}
\def\M{{\mathcal M}}
\def\bM{{\overline \M}}
\def\L{{\mathcal L}}
\def\sM{{\sf M}}
\def\gst{\mathrm{g}_{\mathrm{st}}}
\def\gstt{\widetilde{\mathrm{g}}_{\mathrm{st}}}

\def\be{\begin{equation}}
\def\ee{\end{equation}}

\tableofcontents

\section{Introduction}\label{intro}

JT gravity is a simple model of a real scalar field $\phi$ coupled to gravity in two dimensions \cite{J,T}.   For the case of negative cosmological constant,
the  bulk action in Euclidean signature is
\be\label{JTaction}I_b=  -\frac{1}{2}\int \d^2x \sqrt g\phi(R+2). \ee
Here $R$ is the scalar curvature of the metric tensor $g$.    The model has been fruitfully studied in recent years; see for
example \cite{AP,MSY,EMV,HJ,SSS}.

The particular action of eqn. (\ref{JTaction}) has been chosen to make a model that is as simple as possible.   
It is of interest to consider deforming
the model to a more general class of models.  
What is the natural class of models to consider?   
If we limit ourselves to action functions with at most two derivatives, the most general possibility for the bulk
action is
\be\label{genaction}I_b= -\frac{1}{2}\int\d^2x \sqrt g \left(U(\phi_0) R + 
V(\phi_0) g^{\alpha\beta}\partial_\alpha\phi_0 \partial_\beta \phi_0+W(\phi_0)\right),\ee
with three functions $U,V,W$ of a scalar field $\phi_0$.
   However \cite{Banks,GGG,Ikeda}, it is possible to eliminate two of the three functions by reparametrizing $\phi_0$ and making a Weyl
transformation of the metric.     In fact, if there is a value of $\phi_0$ at which $U'(\phi_0)=0$, 
then the kinetic energy of the fields $\phi_0,g$ is degenerate (even after
gauge-fixing) in expanding around that point, and the theory becomes ill-behaved.   
So we restrict to the case that $U'(\phi_0)$ is everywhere non-zero.  But in this
case, we can just introduce a new scalar field $\phi=U(\phi_0)$.    After then making a Weyl transformation to set $V=0$, we reduce to
\be\label{redto}I_b=-\frac{1}{2}\int\d^2x \sqrt g\left(\phi R + W(\phi)\right). \ee
(This bulk action has to be supplemented by a Gibbons-Hawking-York surface term 
and one usually also wishes to add an Einstein-Hilbert term.   These
will be incorporated in section \ref{thermof}.)   
As this is a Euclidean action, $W$ is the negative of the usual potential energy function.

We want to place a mild restriction on the function $W(\phi)$ so that the behavior near spatial infinity 
-- where the dual quantum mechanical system or random matrix
model lives -- will be the same as in the JT case.   JT gravity is the special case $W(\phi)=2\phi$, and in 
 JT gravity, $\phi\to +\infty$ near spatial infinity.  So the condition that we want is that
$W(\phi)\sim 2\phi$ for $\phi\to+\infty$.    We will refine this condition slightly in section \ref{thermof}.

The classical solutions and black hole thermodynamics of these models, parametrized as in eqn. (\ref{redto}),
have been studied in \cite{GKL,Medved,Cav,Centaur}, among other papers.   
Classical solutions and thermodynamics of a number of dilaton gravity models were studied earlier in a different
parametrization; see for example \cite{MSW,NP}.   See also \cite{VV} for a recent study of the Wheeler-de Witt equation
in the family of models (\ref{redto}).

The present article is devoted to a somewhat more detailed study of these models at the classical level.   In a companion article, it will be argued
that models in this class can be understood as matrix models, generalizing the result of \cite{SSS} for JT gravity.

In section \ref{thermof} of this article, we review the classical solutions of these models and the associated 
thermodynamics.   In section \ref{stability},
we analyze the thermodynamic stability of these black hole solutions. The main result is to show that if
there is a black hole solution at temperature $T$ with negative specific heat, then there is another black hole
solution at the same temperature with positive specific heat and lower free energy.  In section \ref{potentials}, 
we consider phase transitions in these models.   When the temperature is increased, keeping $W$ fixed,  the energy and entropy always
increase monotonically, but not necessarily continuously; there can be first order phase transitions, analogous to the 
Hawking-Page transition \cite{HP,CRJM,LRR}.
In addition, at zero temperature, when $W$ is varied, the ground state can change discontinuously at a first order phase transition.
All of these statements reflect standard ideas about  black hole thermodynamics; see for example \cite{Wald}.     
Finally, in section \ref{closed}, we describe compact smooth Euclidean solutions of models in this class.   
These solutions are de Sitter-like, even though they can arise in models that have black
hole solutions that are asymptotic to Anti de Sitter space.   

\section{Classical Solutions and Thermodynamics}\label{thermof}

In Euclidean signature, a general black hole solution of a theory in the class (\ref{genaction}) can be put in the form\footnote{In fact,
every classical solution of such a theory has a Killing vector field $V^I=\varepsilon^{IJ}\partial_J\phi$, where $\varepsilon^{IJ}$
is the Levi-Civita tensor \cite{GKL}.  Upon picking coordinates $r,t$ so that the Killing vector field generates translations of $t$,
and then shifting $r$ by a suitable function of $t$, one puts the  solution in the form stated in the text.}
\be\label{kiggo} \d s^2=A(r)\d t^2+\frac{1}{G(r)}\d r^2,     ~~~~ \phi=\phi(r), \ee
for some functions $A(r)$, $G(r)$, $\phi(r)$.    In fact, this form is still invariant under reparametrizations 
$r\to \t r(r)$, which can be used to set $G=A$.
However, it is most illuminating to do this only after deducing the equation of motion for $G$. 
 In a black hole solution, $t$ will be a periodic variable with
some period $\beta$, to be determined, so the bulk action can be written as an integral over $r$ only:
\be\label{iggo}I_b=-\frac{\beta}{2} \int \d r \left(-\phi\frac{\d}{\d r} \left(\sqrt{\frac{G}{A}}\frac{\d A}{\d r}\right) + 
\sqrt{\frac{A}{G}} W(\phi)\right). \ee
Hence the equation of motion for $G$ is
\be\label{bobo} \frac{\phi' A'}{\sqrt{AG}} -\frac{\sqrt A}{G^{3/2}} W(\phi)=0. \ee
Having deduced this equation, one can reparametrize the $r$ coordinate to set $G=A$, upon which the  line element becomes
\be\label{higgo}\d s^2 =A(r)\d t^2+\frac{1}{A(r)}\d r^2,\ee
the bulk action simplifies to 
\be\label{liggo}I_b=-\frac{\beta}{2}\int \d r\left(-\phi A'' +W(\phi)\right),    \ee
and the equation (\ref{bobo}) simplifies to
\be\label{obo} \phi' A'- W(\phi)=0. \ee
The equation of motion for $A$ is
\be\label{pobo} \phi''=0,\ee
so $\phi=ar+b$ for constants $a,b$.    The form (\ref{higgo}) of the metric is invariant under 
$(t,r,A)\to (t/a', a'r+b',(a')^2A)$,  and we can fix this remaining
freedom by specifying that
\be\label{zobo}\phi=r. \ee
The equation (\ref{obo}) then simplifies to $A'=W(r)$, so
\be\label{jobo}A(r)=\int_{r_h}^r \d r' W(r'), \ee
for some $r_h$.   These formulas were obtained in \cite{GKL}.    Note that we have found the general solution -- 
in terms of one arbitrary constant $r_h$ -- without
using the Euler-Lagrange equation that comes by varying $\phi$.   This equation in fact does not give additional information.  

Clearly, $A(r_h)=0$, so we interpret $r=r_h$ as the black hole horizon.  The value of $\phi$ at the horizon is therefore
\be\label{woddo}\phi_h=r_h. \ee
  Near $r=r_h$, we have
\be\label{gobi} A(r)=W(r_h)(r-r_h)+\O((r-r_h)^2). \ee
Hence the solution only has the expected Euclidean signature if $W(r_h)\geq 0$.  
More generally, a black hole solution with $\phi=\phi_h$ exists if and only if
$A(r)$, defined in eqn. (\ref{jobo}), is positive for all $r>r_h$.  In other words, the condition is that
\be\label{mod} \int_{\phi_h}^\phi\d \phi' \,W(\phi')>0~~{\mathrm{for~all}}~\phi>\phi_h.\ee

 \def\GHY{{\mathrm{GHY}}}
 \def\EH{{\mathrm{EH}}}
Setting $y=\sqrt{r-r_h}=\sqrt{r-\phi_h}$, we find that the line element near $y=0$ is
\be\label{obi} \d s^2 =\frac{4}{W(\phi_h)}\left(\d y^2+\frac{W(\phi_h)^2}{4} y^2\d t^2\right). \ee
This is smooth at $y=0$ if and only if $t$ has period $4\pi/W(\phi_h)$.   This determines the temperature:
\be\label{zobi}\beta=\frac{4\pi}{W(\phi_h)},~~~ T=\frac{W(\phi_h)}{4\pi}. \ee

As explained in the introduction, we want $W(\phi)\sim 2\phi$ for $\phi\to+\infty$ so as to get a theory that reduces to JT gravity for $r\to\infty$.
More specifically, we will assume that for large $\phi$, $|W(\phi)-2\phi|<1/\phi^{1+\epsilon}$.    In this case, the behavior of $A(r)$ for large $r$ is
\be\label{kini} A(r)=r^2-b +\O(r^{-\epsilon}), \ee
where $b$ is a constant that we will see is a multiple of the energy.    For JT gravity ($W(\phi)=2\phi$ exactly, not just asymptotically), 
the formula $A(r)=r^2-b$ is exact.  In general, a solution with the behavior of eqn. (\ref{kini}) is 
 asymptotic at infinity to an asymptotically $\mathrm{AdS}_2$ solution of JT
gravity.   To keep things simple, we will also assume that $\lim_{\phi\to\infty}\,W'(\phi)=2$, thus excluding small-scale
oscillations in $W(\phi)$ at large $\phi$.

From eqn. (\ref{jobo}) and the definition of $b$, we see that under a change in $\phi_h=r_h$,
\be\label{cobo} \d b = W(\phi_h) \d \phi_h=4\pi T \d \phi_h. \ee
It is natural to guess that as in conventional black hole thermodynamics,
 this will be the first law of thermodynamics $\d E=T\d S$, with $\phi_h$ a linear function of the entropy $S$ and $b$ a linear function of the energy $E$.
To confirm this interpretation, we will evaluate the action for the solution.    This action is interpreted as a classical approximation to $-\log Z= \beta F
=\beta E-S$, where $Z$ is the partition function and $F=E-TS$ is the free energy.
To evaluate the action, as in \cite{MSY},
we put a cutoff on $r$, at some very large value $r=r_\infty$,  and we include a Gibbons-Hawking-York surface term in the action:
\be\label{riffo} I_{\GHY} =-\int_{r=r_\infty} \d t \sqrt h \phi (K-1) =-\beta \sqrt{A(r_\infty)}(K-1). \ee
Here $h=A(r_\infty)$ is the induced metric of the boundary, and $K$ is the extrinsic curvature of the boundary, explicitly $K=A'(r_\infty)/2\sqrt{A(r_\infty)}$.
With $\phi(r)=r$ and $A(r)\sim r^2-b$, we find that the surface term in the action is
\be\label{ziffo} I_\GHY =-\frac{\beta b}{2}. \ee
Remembering that $\phi=r$ and $W=A'$, the bulk action is
\begin{align}\label{giffo}I_b&=-\frac{\beta}{2} \int_{r_h}^{r_\infty}\d r \left(-r A''+A'\right)=-\frac{\beta}{2}\int_{r_h}^{r_\infty} \d r\left(\frac{\d}{\d r}\left(- r A'+2A\right)\right)\cr
&=-\frac{\beta}{2}\left[ -r A'+2A  \right]_{r_h}^{r_\infty}=-\frac{\beta}{2}\left(-2b+r_h W(r_h)\right) =\beta b - 2\pi \phi_h. \end{align}
The total action, including also an additive constant $-S_0$ from a possible Einstein-Hilbert term $I_{\EH}$ in the action, is thus
\be\label{bozo}I=I_b+I_{\GHY}+I_{\EH} = \frac{\beta b}{2} -2\pi \phi_h -S_0.\ee
Setting this equal to $\beta E-S$, we get
\be\label{wozo} E=\frac{b}{2},   ~~~~ S=2\pi \phi_h +S_0, \ee
and we see that eqn. (\ref{cobo}) can indeed be interpreted as the first law.  
The formula $E=b/2$ is the analog for this type of model of the fact that in ordinary gravity in four spacetime dimensions,
the ADM mass is defined in terms of the leading correction to the asymptotic form of the metric at infinity.   For a systematic
framework for deriving the first law, see \cite{IyerWald}.

Based on these results, we can refine our assumptions about the function $W(\phi)$.  
If this function is positive-definite, the condition (\ref{mod}) is satisfied for any $\phi_h$ and therefore there is a black hole
solution for any $\phi_h$.      In this case, since $\phi_h$ can be arbitrarily negative, the black hole entropy is unbounded
below.   This is presumably unphysical.  Note that if and only if the entropy is bounded below, we can pick $S_0$ so that the bound is nonnegative, as one may
expect physically.
(For $W(\phi)$ positive-definite, the  black hole energy will also be unbounded below unless $W(\phi)$ vanishes sufficiently
rapidly for $\phi\to -\infty$.)
So we assume that $W(\phi)$ is negative at least in some range of $\phi$. More specifically, to make the entropy  bounded below,
we want to constrain $W$  so that the set of all $\phi_h$ that satisfy condition (\ref{mod}) is bounded below.
This actually does not imply that $W(\phi)<0$ for $\phi\to-\infty$, but it does imply that $W(\phi)<0$ for some range of $\phi$ and that 
\be\label{nugu}\liminf_{\phi\to-\infty}\,W(\phi)\leq 0.\ee
In other words, for any $\epsilon>0$, there are arbitrarily negative values of $\phi$ with $W(\phi)<\epsilon$.  If instead $W(\phi)$ is bounded below by $\epsilon>0$
for sufficiently negative $\phi$, then  black hole solutions can have arbitrarily negative $\phi_h$ and the entropy and energy are both  unbounded below.

With the result (\ref{wozo}) for the energy, and the fact that $A(r_\infty)$ can be approximated as 
$r_\infty^2-b$ for large $r_\infty$, we find that
the energy of a black hole with given $\phi_h$ can be written as
\be\label{nbn}E(\phi_h)=\frac{1}{2}r_\infty^2-\frac{1}{2}\int_{\phi_h}^{r_\infty} \d \phi\, W(\phi). \ee
It is convenient to subtract the values of $E$ for two different values of $\phi_h$ so as to get a formula that does 
not make reference to the cutoff $r_\infty$:
\be\label{pbn} E(\phi_1)-E(\phi_0)=\frac{1}{2} \int_{\phi_0}^{\phi_1} \d \phi \, W(\phi). \ee   This result can also be found by 
integrating the first law
$\d E=T\d S$.
Suppose that $\phi_0<\phi_1$ and that there is a black hole with $\phi_h=\phi_0$.  Then eqn. (\ref{mod}) with $\phi=\phi_1$ tells us that
\be\label{hbn} E(\phi_1)-E(\phi_0)>0. \ee
This is true whether or not there is a black hole with $\phi_h=\phi_1$. (There may not be one, since eqn. (\ref{mod}) may not hold for 
$\phi_h=\phi_1$.)     Specializing to the case that there is such a black hole, we learn  that black hole
energy is always an increasing function of $\phi_h$.   Of course, black hole entropy is also an increasing function of $\phi_h$,
since the entropy is a multiple of $\phi_h$.

\section{Thermodynamic Stability}\label{stability}

There is a black hole solution at any value of $\phi_h$ such that eqn. (\ref{mod}) is satisfied.    But these solutions are not all thermodynamically
stable.

A basic thermodynamic inequality says that in any thermal ensemble, the heat capacity (or specific heat) is positive:
\be\label{numquat}\frac{\d E}{\d T}>0. \ee
Let us see what this condition means for our black holes.

If there is a black hole with $\phi=\phi_h$, then $W(\phi_h)>0$, or else eqn. (\ref{mod}) is not satisfied for $\phi$ slightly greater than $\phi_h$.
This means that we always have $\d E/\d \phi_h>0$.    On the other hand, since $T(\phi_h)=W(\phi_h)/4\pi$, we have $\d T/\d \phi_h=W'(\phi_h)/4\pi$.
If $W'(\phi_h)>0$, then $\d E/\d\phi_h$ and $\d T/\d \phi_h$ are both positive, and therefore $\d E/\d T>0$.
But if $W'(\phi_h)<0$, then $\d E/\d \phi_h$ and $\d T/\d \phi_h$ have opposite signs and hence $\d E/\d T<0$, violating the laws
of thermodynamics.

The resolution of this point is that a black hole with $W'(\phi_h)<0$ is always thermodynamically unstable
 in the canonical ensemble (defined by specifying the temperature).
In exploring this question, we assume that the condition of eqn. (\ref{mod}) is
satisfied at $\phi_h=\phi_1$, so that there is a black hole solution with that value of $\phi_h$. In particular $W(\phi_1)>0$.     If $W'(\phi_1)<0$,  then, since we assume that $W(\phi)$ is asymptotically increasing (as $2\phi$)  for large $\phi$, there is always some $\phi_2>\phi_1$ with
$W(\phi_2)=W(\phi_1)$.   There may be multiple values $\phi_2>\phi_1$ that satisfy this condition.    Eqn. (\ref{mod}) is not necessarily satisfied
for each possible choice of $\phi_2$, but it is always satisfied at the largest such choice, which we will denote as $\h\phi$.   (Since $W(\phi)$ grows
asymptotically for large $\phi$, we have $W(\phi)>W(\h\phi)>0$ for $\phi>\h\phi$; hence eqn. (\ref{mod}) is trivially satisfied
for $\phi_h=\h\phi$.) 
Moreover, because of eqn. (\ref{nugu}), there is always $\phi_0<\phi_1$ with
$W(\phi_0)=W(\phi_1)$.   Again, if we pick the largest such $\phi_0$, then $W(\phi)>W(\phi_1)>0$ for $\phi_0<\phi<\phi_1$, so if eqn. (\ref{mod}) is satisfied
at $\phi_h=\phi_1$, then it is satisfied at $\phi_h=\phi_0$ and there is a black hole with $\phi_h=\phi_0$.
    So if there is a black hole solution at $\phi_h=\phi_1$ with $W'(\phi_1)<0$, then there always are at least
two more black hole solutions at the same temperature, at least one with $\phi_h>\phi_1$ and one with $\phi_h<\phi_1$.

\begin{figure}
 \begin{center}
   \includegraphics[width=4in]{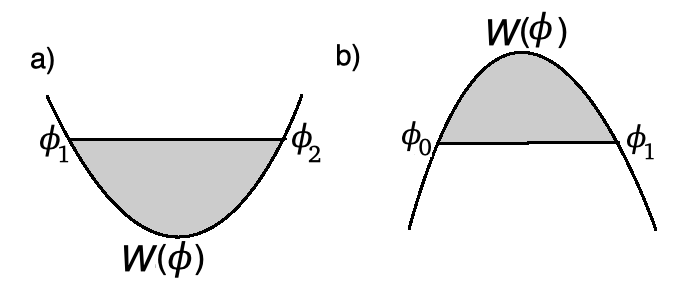}
 \end{center}
\caption{\small Here we assume that there is a black hole at $\phi_h=\phi_1$ with $W'(\phi_1)=0$ and we explore its themodynamic stability.  (a)  In this example, 
there is some $\phi_2>\phi_1$ with 
 with $W(\phi_1)=W(\phi_2)$ and $W(\phi)<W(\phi_1)$ for $\phi_1<\phi<\phi_2$.  A black hole with $\phi_h=\phi_1$ is then 
always thermodynamically disfavored compared to a black hole at $\phi_h=\phi_2$ (if such a black hole exists).    They have the same temperature, but the one with
$\phi_h=\phi_2$ has lower free energy.   The free energy difference between the two black holes is the area of the shaded region (times $2\pi$).
(b)   In this example, there is $\phi_0<\phi_1$ with $W(\phi_0)=W(\phi_1)$ and $W(\phi)>W(\phi_1)$ for $\phi_0<\phi<\phi_1$.   With our assumptions,
such a $\phi_0$ always exists and there is  a black hole with $\phi_h=\phi_0$.   The black hole at $\phi_h=\phi_0$ is always thermodynamically
favored compared to the one at $\phi_h=\phi_1$; the free energy difference between them is determined by the area of the shaded region. \label{example}}
\end{figure}

To determine which solution is thermodynamically dominant, we need to compare their free energies.  For definiteness, we first write formulas for the
case $\phi_2>\phi_1$.  
Let $T=W(\phi_1)/4\pi=W(\phi_2)/4\pi$ be the  temperature of the black holes at $\phi_h=\phi_1$ or $\phi_h=\phi_2$.  
It is convenient to also define $T(\phi)=W(\phi)/4\pi$ (this is a formal definition and $T(\phi)$ is not really a $\phi$-dependent temperature).
The entropy difference between the two  black holes  is
\be\label{hbu}\Delta S=2\pi (\phi_2-\phi_1). \ee
The energy difference from eqn. (\ref{pbn}) is
\be\label{nubu}\Delta E=\frac{1}{2} \int_{\phi_1}^{\phi_2}\d\phi\, W(\phi)=2\pi \int_{\phi_1}^{\phi_2}\d\phi\, T(\phi), \ee
The free energy difference between the black hole at $\phi_h=\phi_2$ and the black hole at $\phi_h=\phi_1$ is
\be\label{zubu}  \Delta F=\Delta E-T\Delta S =2\pi\int_{\phi_1}^{\phi_2} \d\phi \left(T(\phi)-T\right). \ee
In other words, $\Delta F/2\pi$ is the difference between the area under the curve $y=T(\phi)$ and the area under the straight line $y=T$ over the interval $[\phi_1,\phi_2]$.
If $\phi_2$ is the smallest solution of $W(\phi_2)=W(\phi_1)$ in the region $[\phi_1,\infty)$, then as depicted in fig. \ref{example}(a), we have $W(\phi)<W(\phi_1)$ for $\phi_1<\phi<\phi_2$.   It then follows from eqn. (\ref{zubu}) that  a black hole with $\phi_h=\phi_2$, 
if it exists, has lower free energy than the black hole with $\phi_h=\phi_1$.  The free energy difference between them is determined by
 the area of the shaded region in the figure.
  This is actually not enough to show that the black hole at $\phi_h=\phi_1$
is unstable because there may be no black hole solution with $\phi_h=\phi_2$. (Even if eqn. (\ref{mod}) is satisfied at $\phi_h=\phi_1$, it may not be satisfied at $\phi_h
=\phi_2$.)   To complete the argument that there is a black hole with $\phi_h>\phi_1$ that is more stable than the black hole at $\phi_h=\phi_1$, we have to
look at all the solutions of $W(\phi)=W(\phi_1)$ for $\phi>\phi_1$, and show that at least one of them is associated to a black hole that is thermodynamically favored
over the black hole with $\phi_h=\phi_1$.

We postpone this for a moment and consider black holes with $\phi_h<\phi_1$.    Let $\phi_0$ be the largest solution of $W(\phi)=W(\phi_1)$ for $\phi<\phi_1$.
We have already explained that there is a black hole with $\phi_h=\phi_0$.
This black hole turns  out to be  thermodynamically favored over the black hole
at $\phi_h=\phi_1$.   The free energy difference between the black hole at $\phi_h=\phi_1$ and the one at $\phi_h=\phi_0$ is just given by eqn. (\ref{zubu}),
with $\phi_1$ and $\phi_2$ replaced by $\phi_0$ and $\phi_1$:
\be\label{zzubu}  \Delta F=2\pi\int_{\phi_0}^{\phi_1} \d\phi \left(T(\phi)-T\right). \ee
But this is now positive, because  
 $T(\phi)>T$ in the integration region.
Indeed
(fig. \ref{example}(b)),
 $W(\phi)>W(\phi_1)$ and hence $T(\phi)>T=T(\phi_1)$  for $\phi\in [\phi_0,\phi_1]$.
    Thus the black hole at $\phi_h=\phi_1$ has greater free energy than the one
at $\phi_h=\phi_0$.   The free energy difference between them is determined by the area of the shaded region in fig. \ref{example}(b).

 \begin{figure}
 \begin{center}
   \includegraphics[width=4in]{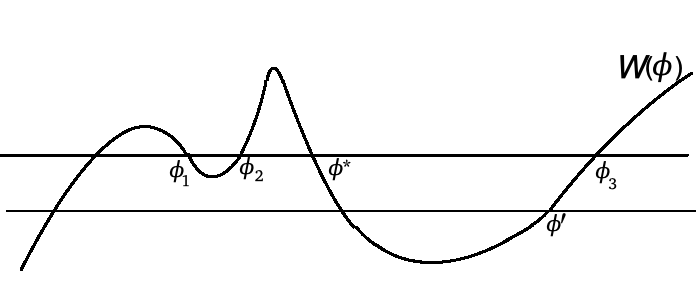}
 \end{center}
\caption{\small  This figure depicts a function $W(\phi)$ with the properties needed to illustrate an argument in the text.   The lower horizontal line
intersects points with  $W=0$ and the upper one intersects points with
$W=W(\phi_1)>0$.   We have $W(\phi_1)=W(\phi_2)=W(\phi^*)=W(\phi_3)$ with $\phi_1<\phi_2<\phi^*<\phi_3$;
moreover $W'(\phi_1), W'(\phi^*)<0$ while $W'(\phi_2),W'(\phi_3)>0$.   
For simplicity, we have assumed that the region  in which $\phi>\phi_1$ and $W(\phi)<0$ is a connected interval with upper end-point $\phi'$.
This ensures that for $\phi_h$ equal to $\phi_1$ or $\phi_2$, the condition (\ref{mod}) for existence of 
a black hole solution with given $\phi_h$ only has to be checked
for $\phi=\phi'$.       If $W$ is such
that $\int_{\phi_1}^{\phi'}\d \phi \,W(\phi)>0$ but $\int_{\phi_2}^{\phi'}\d\phi, W(\phi)<0$, then there is a black hole with $\phi_h=\phi_1$ but
none with $\phi_h=\phi_2$.   As explained in the text, in this case the black hole at $\phi_h=\phi_3$ is thermodynamically favored over the one
at $\phi_h=\phi_1$.   \label{Messy}}
\end{figure}

This shows that a black hole with $W'(\phi_h)<0$ is thermodynamically disfavored, but we would still like to understand better the role of black holes
with $\phi_h>\phi_1$.   Since $W'(\phi_1)<0$ but $W(\phi)>W(\phi_1)$ for sufficiently large $\phi$, the number of solutions of $W(\phi)=W(\phi_1)$
in the interval $[\phi_1,\infty)$ is odd.   If there is only one such solution $\phi_2$, then $W(\phi)>W(\phi_2)>0$ for all $\phi>\phi_2$, so eqn. (\ref{mod})
is satisfied at $\phi_h=\phi_2$ and there is a black hole with $\phi_h=\phi_2$.   As we have already shown,
this black hole has lower free energy than the one at $\phi_h=\phi_1$.   Let us look at the next case that there are three solutions of $W(\phi)=W(\phi_1)$
for $\phi\in[\phi_1,\infty)$.   Two of them will have $W'(\phi)>0$ and one has $W'(\phi)<0$.    Let the solutions with $W'(\phi)>0$ be at $\phi=\phi_2,\phi_3$ with $\phi_2<\phi_3$, and write $\phi^*$ for the third solution with $W'(\phi)<0$.    A potential with these properties is sketched in fig. \ref{Messy}.

We already know that if there is a black hole at $\phi_h=\phi_2$,
then it is thermodynamically favored over the one at $\phi=\phi_1$.  However, in general there is no such black hole
because eqn. (\ref{mod}) may not be satisfied at $\phi_h=\phi_2$.  For this to happen, there must be a portion of the interval $[\phi_2,\phi_3]$ with $W(\phi)<0$, as in the
example sketched
in fig. \ref{Messy}.  There is always a black hole with $\phi_h=\phi_3$ (since $W(\phi)>W(\phi_3)>0$ for all $\phi>\phi_3$, ensuring that eqn. (\ref{mod}) is satisfied).  But  in general this black hole may have greater free energy than the one at $\phi_h=\phi_1$.   We want to show
 that either there is a black hole with $\phi_h=\phi_2$, and therefore the black hole with $\phi_h=\phi_1$ is thermodynamically disfavored, or
the black hole with $\phi_h=\phi_3$ has lower free energy than the one at $\phi_h=\phi_1$, and again the black hole with $\phi_h=\phi_1$
is themodynamically disfavored.

For eqn. (\ref{mod}) not to be  satisfied at $\phi=\phi_h$ means that there is some $\phi'\in [\phi_2,\phi_3]$ with
\be\label{ubbu}\int_{\phi_2}^{\phi'} \d \phi \,W(\phi)\leq 0. \ee
Note that necessarily $\phi'>\phi^*$.   
If $\phi$ is contained in either the interval $[\phi_1,\phi_2]$ or the interval $[\phi',\phi_3]$, then 
  $W(\phi)<W(\phi_1)$ and hence $T(\phi)<T$.
Combining these facts,
\begin{align}\label{tubbu} \int_{\phi_1}^{\phi_3} T(\phi)\d \phi & =\int_{\phi_1}^{\phi_2} T(\phi)\d \phi +\int_{\phi_2}^{\phi'}T(\phi)\d\phi +\int_{\phi'}^{\phi_3} T(\phi)\d\phi
\cr &\leq  T((\phi_3-\phi')+(\phi_2-\phi_1))<T(\phi_3-\phi_1). \end{align}  We have used (\ref{ubbu}) for the integral in the interval $[\phi_2,\phi']$, 
and the bound $T(\phi)<T$ for the
integrals in the other two intervals.   But eqn. (\ref{tubbu}), together with eqn. (\ref{zubu}),  precisely says that the free energy of the black hole with $\phi_h=\phi_3$
is less than the one with $\phi_h=\phi_1$.    

If instead there is a black hole with $\phi_h=\phi_2$, then eqn. (\ref{ubbu}) is false and eqn. (\ref{tubbu}) may also be false.  Regardless,
we have learned that as one increases $\phi_h$ from $\phi_1$, the first black hole that one encounters that has the same temperature as the one
at $\phi_h=\phi_1$ has lower free energy than the one at $\phi_h=\phi_1$.  This black hole is at either $\phi_h=\phi_2$ or $\phi_h=\phi_3$, depending on $W$.

A similar statement holds in general.
In general, if $W'(\phi_1)<0$, then
for $\phi>\phi_1$, there might be any number $n\geq 1$ of solutions of $W(\phi)=W(\phi_1)$ with $W'(\phi)>0$.   (There are then $n-1$ solutions
with $W'(\phi)<0$.)    Let us label these solutions 
as $\phi_2<\phi_3<\cdots < \phi_{n+1}$.    There is always a black hole with $\phi_h=\phi_{n+1}$, and for $2\leq k\leq n$,  there may or not be a black hole with
$\phi_h =\phi_k$.  Let $k$ be the smallest element in the set $\{2,3,\cdots, n+1\}$ such that there is a black hole at $\phi_h=\phi_k$.
An argument similar to the one already explained shows that this black hole has lower free energy than the one at $\phi=\phi_1$.   So if $W'(\phi_1)<0$,
 there is always a black
hole with $\phi_h>\phi_1$ that has the same temperature as the one with $\phi_h=\phi_1$ and lower free energy.

We have shown that if there is a black hole with $\d E/\d T<0$, then
there are black holes of the same temperature  but lower free energy both at larger values of $\phi_h$ and at smaller values of $\phi_h$.   Which of these
has the lowest free energy?
 This is part of what we will discuss next.

\section{Potentials and Phase Transitions}\label{potentials}

We now have the tools to get a general picture of the thermodynamics of classical black holes in these models.    

Assuming that the black hole entropy is bounded below, there is a smallest value of $\phi_h$, say $\phi_h=\phi_0$, at which there is a classical black 
hole solution.   This is the smallest value at which eqn. (\ref{mod}) is satisfied.   The black hole at $\phi_h=\phi_0$ has the smallest possible
entropy, and, according to eqn. (\ref{pbn}), it also has the smallest energy of any black hole solution.

Necessarily $W(\phi_0)=0$; otherwise, eqn. (\ref{mod}) is still satisfied for $\phi_h$ slightly less than $\phi_0$.
Eqn. (\ref{mod}) for $\phi$ slightly greater than $\phi_0$ implies that $W(\phi)>0$ in that region, and since $W(\phi_0)=0$, it follows that 
we also have  $W'(\phi)>0$ for $\phi$ slightly greater than $\phi_0$. (Generically $W'(\phi_0)>0$, but this need not always be true.)
    Of course, we also have $W(\phi)>0$ 
and $W'(\phi)>0$ for sufficiently large $\phi$.

We will first assume that $W(\phi)>0$ for all $\phi>\phi_0$.    
In this case, eqn. (\ref{mod}) is satisfied for all $\phi_h\geq\phi_0$, so there is a
black hole solution that is asymptotic to $\mathrm{AdS}_2$ spacetime at spatial infinity for any assumed value $\phi_h\geq\phi_0$. 
Since $W(\phi_0)=0$, the  black hole with $\phi_h=\phi_0$ will be a zero temperature, extremal black hole, with some energy $E_0$.  The temperature is positive for $\phi_h>\phi_0$.   $W(\phi)$ being positive means that the black hole energy is a monotonically increasing function of $\phi_h$.
  
The simplest case is that  $W'(\phi)>0$  in the whole range $[\phi_0,\infty)$.   Then the black hole
temperature, which is a multiple of $W$, is a  monotonically increasing functions of $\phi_h$ throughout the whole range.   There is a unique black hole of any given temperature, and it is always thermodynamically stable.     

\begin{figure}
 \begin{center}
   \includegraphics[width=4in]{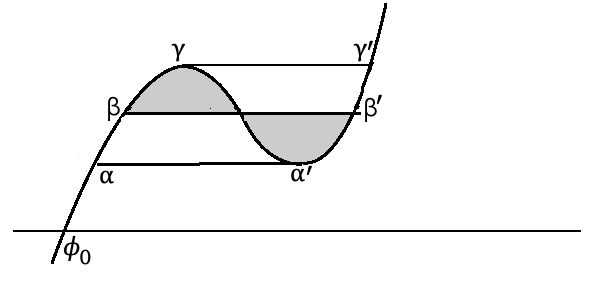} \end{center}
   \caption{\small   A function $W(\phi)$ that is positive for $\phi>\phi_0$ but such that $W'(\phi)$ has two zeroes in that region.  Horizontal lines
   intersect points with the same value of $\phi_h$, corresponding to black holes with the same temperature.
   Starting at $\phi_0$ and moving up the curve, the black hole remains thermodynamically stable up to and beyond
   the point labeled $\alpha$.   Thermodynamic stability is lost at the point $\beta$, which is characterized by the fact
   that the shaded regions above and below the horizontal line $\beta\beta'$ have the same area.   As the black
   hole temperature is increased, the point that labels the black hole solution moves continuously along the curve
   from $\phi_0$ up to $\beta$, jumps to $\beta'$, and then
   continues upwards from there.    The jump from $\beta$ to $\beta'$ is a first order phase transition, analogous to the Hawking-Page phase transition.
  \label{Comparison}}
\end{figure}

Now let us keep the assumption that $W(\phi)>0$ for all $\phi>\phi_0$, but drop the assumption that $W'(\phi)>0$ in that range.    Since $W'(\phi)>0$ for $\phi$
slightly larger than $\phi_0$ and
$W'(\phi)>0$ for very large $\phi$, the function $W'(\phi)$ has an even number of zeroes for $\phi>\phi_0$.  
For example, in fig. \ref{Comparison}, we illustrate a case with two zeroes.    For such a potential,  for a certain range of temperatures,  there are three black hole
solutions, two with $W'(\phi_h)>0$ and one with $W'(\phi_h)<0$.   In fig. \ref{Comparison}, this is true on the portion of the curve between the points labeled
$\alpha$ and $\gamma'$.   

Horizontal lines in the figure connect points with the same value of $W(\phi)$, corresponding to black hole solutions with the same temperature.
For example, the pairs of points labeled $\alpha$ and $\alpha'$ or $\gamma$ and $\gamma'$ represents pairs of black hole solutions with the same temperature.
From eqn. (\ref{zubu}), it follows that the black hole at $\alpha $ is more stable thermodynamically  than the one at $\alpha'$ and but the black hole at $\gamma$
is less stable than the one at $\gamma'$.    As one moves up the curve from $\alpha$, there is a first order phase transition at a point $\beta$ that is characterized
by the fact that the shaded regions above and below the horizontal curve $\beta\beta'$ have the same area.    That condition means that the black holes
corresponding to the points $\beta$ and $\beta'$ have the same free energy.   Of course, the black hole at $\beta'$ has higher energy and entropy than
the one at $\beta$, so it will be will be favored as soon as the temperature is increased further.

Thus, as the temperature is increased from 0, one starts at $\phi_0$, moves up the curve to $\beta$, then jumps to $\beta'$ and proceeds upwards from there.
The jump from $\beta$ to $\beta'$ is a first order phase transition, analogous to the Hawking-Page phase transition.
The energy and entropy both increase discontinuously as a function of the temperature, leaving fixed the free energy.    Because
the energy jumps upwards as the temperature is increased, the heat capacity $\d E/\d T$ has a delta function at the transition point
with a positive coefficient.  This positivity
is consistent with general thermodynamic inequalities.

A more general case in which  $W(\phi)>0$ for all $\phi>\phi_0$, but $W'(\phi)$ has more than two zeroes for $\phi>\phi_0$, can be analyzed similarly.
The only real difference is that in general, as the temperature is increased, the system may go through more than one first order phase transition.

What happens if we drop the assumption that $W(\phi)>0$ for all $\phi>\phi_0$ (but we continue to define $\phi_0$ as the smallest value of $\phi_h$
at which there is a black hole solution)?   The analysis is similar, with the sole difference that there is
no classical black hole solution with a value of $\phi_h$ such that $W(\phi_h)<0$.   In other words, when we assume that
$W(\phi)>0$ for $\phi>\phi_0$, it follows that there is 
a classical black hole solution for all $\phi_h\geq \phi_0$, but as we have seen, there  can be gaps in the  values of $\phi_h$ that correspond to thermodynamically
stable black holes.
When $W(\phi)$ is not assumed to be positive for all $\phi>\phi_0$, there are gaps in the allowed values of $\phi_h$  just at the classical level
because classical black hole
solutions do not exist in regions with $W(\phi_h)<0$.   
These gaps imply the occurrence of first order phase transitions.   Indeed, 
since $\phi_h=\phi_0$ at zero temperature and $\phi_h$ becomes large at high temperatures,
it follows that as the temperature is increased, 
$\phi_h$ must at some point jump over the gaps.  Such jumping represents a first order phase transition.
   In general there may be multiple first order phase
transitions, partly due to classical gaps and partly due to considerations of thermodynamic stability.

 \begin{figure}
 \begin{center}
   \includegraphics[width=3.3in]{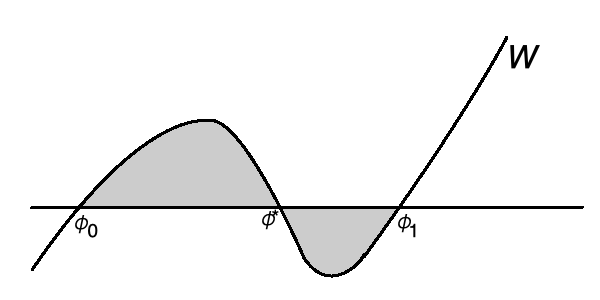}  
 \end{center}
\caption{\small  In this example, the function $W(\phi)$ vanishes precisely at $\phi_0,\phi_1$, and $\phi^*$. It is positive for $\phi>\phi_1$
and negative for $\phi<\phi_0$.   There is always an extremal black hole solution of zero temperature with $\phi_h=\phi_1$.   There is an extremal black hole
solution of zero temperature with $\phi_h=\phi_0$ if and only if the shaded area above the horizontal line exceeds the shaded area below the horizontal line.  (This
is the condition for eqn. (\ref{mod}) to hold at $\phi_h=\phi_0$.) 
If a black hole solution exists at $\phi_h=\phi_0$, it is the true ground state.  If one varies $W(\phi)$ so that the shaded area above
the horizontal line no longer exceeds the shaded area below the horizontal line,
then there is a first order phase transition and the ground state jumps to $\phi_h=\phi_1$.
  \label{Triple}}
\end{figure}

 \begin{figure}
 \begin{center}
   \includegraphics[width=3.3in]{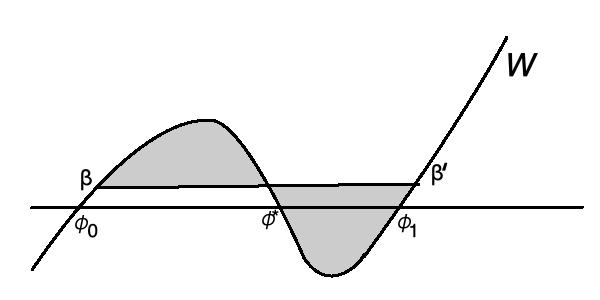}   \end{center}
\caption{\small In the model of fig. \ref{Triple}, the zero temperature ground
   state is at $\phi_h=\phi_0$.    As one increases the temperature, one reaches a first order phase transition at which the
   system jumps across the  ``gap'' that consists of the semi-open interval $[\phi^*,\phi_1)$, where there
   is no classical black hole solution.    The phase transition occurs at a point $\beta$
   characterized by the fact that the shaded regions above and below the horizontal
   line $\beta\beta'$ have the same area.    Thermodynamically stable black holes start at $\phi_0$ at zero temperature, 
   follow the curve from $\phi_0$ to $\beta$ as the temperature is
   increased, jump to $\beta'$, and then follow the curve upward as the temperature
   is increased further.  The interval $(\beta,\beta')$ across which the system jumps contains the interval $[\phi^*,\phi)$ where
   there is no black hole solution, but it is strictly larger: it also contains intervals in which classical black hole solutions
   exist but are not thermodynamically stable.
  \label{Triplets}}
\end{figure}

We illustrate this with an example in fig. \ref{Triple}.   In the figure, $W(\phi)$ is assumed to vanish precisely at three points $\phi_0<\phi^*<\phi_1$,
with $W'(\phi_0), W'(\phi_1)>0$ and $W'(\phi^*) $ negative.   $W(\phi)$ is positive for $\phi>\phi_1$ and negative for $\phi<\phi_0$.   
We assume that $W$ is such that there is a black hole with $\phi_h=\phi_0$; there always is one with $\phi_h=\phi_1$.
 The black holes with $\phi_h$ equal to $\phi_0$ or $\phi_1$
are both extremal black holes with zero temperature, since $W(\phi_0)=W(\phi_1)=0$.    However, eqn. (\ref{pbn}) tells us  that the black 
hole with $\phi_h=\phi_0$ has lower energy.  So at zero temperature, the stable  black hole is the one at $\phi_h=\phi_0$.

What are the possible values of $\phi_h$ for black holes in this model?   Since equation (\ref{mod}) is satisfied at $\phi_h=\phi_0$, it is also satisfied
for $\phi_h$ sufficiently close to $\phi_0$.    And it is 
trivially satisfied for $\phi\geq \phi_1$, where $W(\phi)>0$.   But there is a gap between the two families of black hole solutions, since there is no black hole
solution with $W(\phi_h)<0$.    
As the temperature is increased from zero, at first the system follows continuously the family of classical black hole solutions that starts at $\phi_h=\phi_0$.
But at a certain temperature, there is a first order phase transition and the system jumps across the gap to the region $[\phi_1,\infty)$.    This is sketched
in fig. \ref{Triplets}.   The transition occurs at a point $\beta$ characterized by the fact that the shaded regions above and below the horizontal line $\beta\beta'$
have equal area.

Clearly, the qualitative picture is the same regardless of what mechanism produces a gap in the range of values of $\phi_h$
for a thermodynamically stable black hole.

The general picture is as follows.   In the half-line $[\phi_0,\infty)$, there are in general open intervals that represent values of $\phi$
that are not horizon values of a thermodynamically stable black hole.   These gaps are present because there is no classical
black hole  solution with $W(\phi_h)<0$, because a classical black hole solution with $W'(\phi_h)<0$ is always thermodynamically
unstable, and because black holes with some values of $\phi_h$ are thermodynamically unstable even though $W(\phi_h)$ and
$W'(\phi_h)$ are both positive.  At zero temperature, the true ground state is at $\phi_h=\phi_0$.   As the temperature
is increased from 0 to infinity, the energy and entropy of the black hole, and the horizon value $\phi_h$, increase monotonically
but not necessarily continuously. At a value of $\phi_h$ that is in the interior of the set of allowed and thermodynamically stable values, we always
have $W'(\phi_h)>0$ and hence the energy and entropy are smoothly increasing functions of the temperature.
  Whenever $\phi_h$ reaches a gap in the spectrum of allowed and thermodynamically stable values, 
there is a first order phase transition and $\phi_h$ jumps
across the gap, with a discontinuous increase in the energy and entropy.

The phase transitions that we have just described occur, for suitable $W(\phi)$, when the temperature is varied keeping $W(\phi)$ fixed.
They are analogs of the Hawking-Page transition in higher dimensions  \cite{HP,CRJM,LRR}.     There are also phase transitions at zero
temperature when $W(\phi)$ is varied.   For example, in fig. \ref{Triple}, if one varies $W(\phi)$ so that the shaded area above the horizontal
line no longer exceeds the shaded area below the horizontal line, then there is no longer a black hole solution with $\phi_h=\phi_0$, and
the ground state jumps to $\phi_h=\phi_1$.   This is a first order phase transition; the ground state energy (which is the same
as the free energy since the temperature is zero) varies continuously but not smoothly as a function of $W$, and there is a 
discontinuity in the ground state entropy.

\section{Closed Universes}\label{closed}

Let us return now to the classical equations that were discussed in section \ref{thermof}.   Consider a classical solution such that $A=0$ at a point at which $\phi=\phi_0$ and assume that $W(\phi_0)>0$.
If eqn. (\ref{mod}) is satisfied for $\phi_h=\phi_0$, then this solution describes a classical black hole that is asymptotic to $\mathrm{AdS}_2$ at infinity.
If not, there  is some $\phi_1>\phi_0$ with
\be\label{lono}\int_{\phi_0}^{\phi_1}\d\phi\, W(\phi)=0.   \ee
Pick $\phi_1$ to be as small as possible satisfying this condition.

In this situation, we are not going to get a spacetime asymptotic to $\mathrm{AdS}_2$ at infinity, because the radial coordinate $r$ will run  over the compact
range $\phi_0\leq r\leq \phi_1$.    The line element is the familiar
\be\label{ono}\d s^2=A(r)\d t^2+\frac{1}{A(r)}\d r^2, \ee
with $A(\phi_0)=A(\phi_1)=0$ and $A(r)>0$ for $\phi_0<r<\phi_1$.    If it is possible to compactify the $t$ direction so as to get a smooth manifold
with this line element, then this line element will describe a compact Euclidean signature solution of the theory.
As discussed in section \ref{thermof}, to avoid a conical singularity at $r=\phi_0$, we must take $t$ to be a periodic variable with period
$\beta= 4\pi/W(\phi_0)$.    But now we have to apply a similar logic at the other end; to avoid a conical singularity at $r=\phi_1$,
$t$ must be a periodic variable with period $\beta=-4\pi/W(\phi_1)$.  (The reason for the minus sign is that $\phi_0$ is a minimum of $r$ but
$\phi_1$ is a maximum; one can exchange maxima and minima of $r$ by changing the sign of $r$, but because $\d A/\d r=W$, this is equivalent
to changing the sign of $W$.)       The condition that makes it possible to simultaneously avoid singularities
at both ends is simply
\be\label{gono}W(\phi_0)=-W(\phi_1).   \ee
If this condition is obeyed, we get a smooth Euclidean solution that is topologically a two-sphere.

In other words, to find compact Euclidean solutions, we must adjust the two variables $\phi_0$ and $\phi_1$ to satisfy the two equations (\ref{lono}) and (\ref{gono}).
In addition,  to make $A$ positive for $\phi_0<r<\phi_1$, $W$ must be such that 
\be\label{weggo} \int_{\phi_0}^\phi\d\phi'\, W(\phi')  >0~{\mathrm {for}}~
\phi_0<\phi<\phi_1.\ee   We also need $W(\phi_0)>0$ to keep $\beta $ finite.

Since we have two equations for two unknowns, no special fine-tuning is needed to satisfy these conditions.   
For a generic $W$, one would expect solutions
to be isolated and nondegenerate.\footnote{Nondegeneracy means that  if eqn. (\ref{lono}) and (\ref{gono}) are satisfied for some pair $\phi_0,\phi_1$,   then in expanding around this solution, there are no zero-modes.   In other words,
a solution is nondegenerate if in perturbing around this solution,  the equations
cannot be satisfied to first order in the perturbation.}   Isolated, nondegenerate solutions are stable and generic in the sense that if  for some $W$
there is such a solution for some pair $\phi_0,\phi_1$, then for any sufficiently nearby $W$ there is an isolated, nondegenerate solution for nearby values of $\phi_0,\phi_1$.

If one does not worry about nondegeneracy, it is quite easy to find examples of $W$ for which solutions exist.   
For instance, pick any potential with $W(\phi)=-W(-\phi)$ and $W'(0)<0$.   (Note that JT gravity with negative cosmological constant
has $W(\phi)=2\phi$, which satisfies the first
condition but not the second.)    If $\phi_1=-\phi_0$, then the pair $\phi_0,\phi_1$ certainly satisfies eqns.
(\ref{lono}) and (\ref{gono}).   Because $W'(0)<0$, if $\phi_0$ is sufficiently small and negative then the additional condition
(\ref{weggo})
is  also satisfied.  Since there is a whole range of allowed values of $\phi_0$, these solutions are certainly
not isolated and nondegenerate.   A generic small perturbation of $W$ will remove this degeneracy, and a suitable small perturbation would leave 
 us with a finite set of isolated,
nondegenerate solutions.

  JT gravity with positive cosmological
constant can be described by $W(\phi)=-2\phi$, with $W'(\phi)<0$ (see \cite{MM}).   Since the solutions described in the last paragraph are supported in a region
with $W'(\phi)<0$, they are qualitatively similar to solutions of JT gravity with positive cosmological constant, even though they can arise in a model
that has the asymptotically $\mathrm{AdS}_2$ black hole solutions that we have studied in the present paper.    This is reminiscent of the embedding
of a portion of de Sitter space  in a world that is asymptotic to $\mathrm{AdS}_2$ in the centaur geometry \cite{Centaur}.  
The centaur geometry actually motivated a previous discussion of compact Euclidean solutions of 
dilaton gravity.\footnote{See Case (ii) in Appendix D of \cite{AGH}.   The behavior of $W(\phi)$ for
$\phi\to\pm\infty$ assumed in that paper is different from our assumptions in the present paper, but this is not very important for
compact Euclidean solutions, which only probe a finite range of $\phi$.}

The compact Euclidean solutions described here can be continued to Lorentz signature by $t\to \i t$.  After this continuation,
the solution describes a Lorentz signature manifold in $1+1$ dimensions, with a spatial slice that is a circle.  (One can
also take a universal cover of the spatial slice, to get a solution that has a spatial slice of infinite length with periodic initial data.)  
One is tempted to think that this solution describes a pair of black holes at antipodal points on the circle (or a periodic array of
black holes, after passing to the universal cover), but this is actually oversimplified. 
Penrose diagrams in two-dimensional dilaton gravity can be rather complicated.  For example, fig. 3 of \cite{MNY}
illustrates some of what can happen.

\noindent{\it Acknowledgment}   Research  supported in part by  NSF Grant PHY-1911298.

\bibliographystyle{unsrt}

\end{document}